\documentclass[lettersize,journal]{IEEEtran}
\usepackage{amsmath,amsfonts}
\usepackage{algorithmic}
\usepackage{algorithm}
\usepackage{array}
\usepackage[caption=false,font=normalsize,labelfont=sf,textfont=sf]{subfig}
\usepackage{textcomp}
\usepackage{stfloats}
\usepackage{url}
\usepackage{verbatim}
\usepackage{mathtools}
\newcommand{\re}[1]{\mathrm{#1}}
\usepackage{graphicx}
\usepackage{subfig} 
\usepackage{bm}
\usepackage{amssymb}
\usepackage{enumitem}
\usepackage[hidelinks]{hyperref}
\usepackage{graphicx}
\usepackage{multirow} 
\usepackage{makecell}  
\usepackage[justification=centering]{caption}
\usepackage{booktabs}

\usepackage{bm}

\renewcommand\eqref[1]{(\ref{#1})}

\usepackage{cite}
\begin{document}

\title{Intelligent Optimization of Wireless Access Point Deployment for Communication-Based Train Control Systems Using Deep Reinforcement Learning}

\author{Kunyu Wu, Qiushi Zhao, Zihan Feng,~Yunxi~Mu,~Hao~Qin,~Xinyu~Zhang, and Xingqi~Zhang~\IEEEmembership{Senior~Member,~IEEE}
        \thanks{Received XXX; revised XXX. This work was supported in part by the XXX. (Kunyu Wu and Qiushi Zhao are equally contributed to this paper. Corresponding author: Hao Qin.)}
        \thanks{Kunyu Wu, Qiushi Zhao and Zihan Feng are with the School of Electronics and Information Engineering, Sichuan University, Chengdu, 610017, China.}
        \thanks{Hao Qin is with the School of Electrical and Electronic Engineering, University College Dublin, Dublin, D04 V1W8, Ireland, and also with the School of Electronics and Information Engineering, Sichuan University, Chengdu, 610017, China (e-mail: hao.qin@scu.edu.cn).}
	\thanks{Yunxi Mu is with the College of Engineering, Peking University, Beijing, China.}
        \thanks{Xingqi Zhang is with the Department of Electrical and Computer Engineering, University of Alberta, Edmonton, AB T6G 1H9, Canada, and also with the School of Electrical and Electronic Engineering, University College Dublin, Ireland (e-mail: xingqi.zhang@ualberta.ca).}
        \thanks{Xinyue Zhang is with the School of Electrical and Electronic Engineering, University College Dublin, Ireland (e-mail: xinyue.zhang@ucd.ie).}
\thanks{Color versions of one or more of the figures in this paper are available online at http://ieeexplore.ieee.org.}
\thanks{Digital Object Identifier: XXXX}
}

\markboth{IEEE Transactions on Intelligent Transportation Systems}%
{Shell \MakeLowercase{\textit{et al.}}: A Sample Article Using IEEEtran.cls for IEEE Journals}


\maketitle

\begin{abstract}
Urban railway systems increasingly rely on communication-based train control (CBTC) systems, where optimal deployment of access points (AP) in tunnels is critical to ensure robust wireless coverage. Traditional methods, such as empirical model-based optimization algorithms, suffer from excessive measurement requirements and suboptimal solutions, while machine learning (ML) approaches often face limited adaptability in complex tunnel environments. This paper proposes a deep reinforcement learning (DRL)-driven framework that integrates parabolic wave equation (PWE) channel modeling methods, conditional generative adversarial network (cGAN)-based data augmentation, and a dueling deep Q-network (Dueling DQN) for AP placement optimization in tunnels. The PWE method produces high-fidelity path loss distributions for a small subset of AP positions, which are then expanded by the cGAN to generate realistic high-resolution path loss maps for all candidate positions, significantly reducing simulation cost while preserving physical accuracy. In the proposed DRL formulation, the state space captures AP positions and signal coverage, the action space defines discrete AP adjustments, and the reward function incentivizes signal improvement while penalizing deployment costs. The dueling DQN architecture decouples state value and action advantage estimation to enhance convergence speed, improve exploration–exploitation balance, and increase the likelihood of reaching globally optimal configurations. Comparative experiments against a conventional Hooke–Jeeves (HJ) optimizer and a traditional DQN demonstrate that the proposed hybrid method achieves superior performance, delivering AP configurations with higher average received power and better worst-case coverage, while being more computationally efficient. This work bridges high-fidelity electromagnetic simulation, generative modeling, and AI-driven optimization, offering a scalable and data-efficient solution for next-generation CBTC systems in complex tunnel environments.

\end{abstract}


\section{Introduction}

Urban railway systems are increasingly dependent on communication-based train control (CBTC) systems to ensure safety and efficiency~\cite{zhu2014design,qin2023physics, song2019Propagation, Wang2015ACognitive, Qin2025Physics}, where access points (APs) serve as critical bridges between trains and ground infrastructure. In railway tunnels characterized by confined electromagnetic wave-attenuating environments, the optimal placement of APs is paramount to guarantee continuous high-quality wireless coverage for real-time data transmission and control~\cite{Wen2018Access, Saki2020Comprehensive}. 

From a system design perspective, CBTC imposes stringent requirements on communication latency and reliability. This makes exhaustive electromagnetic simulations for all possible AP positions computationally prohibitive, especially when using high-fidelity models such as the parabolic wave equation (PWE)~\cite{Qin23comparative, Guan2016Vector, QIN25TOA, zhao2023embedding, OZGUN20112638, Qin23_AWPL_SSSPE}. Although these simulations capture complex propagation effects with physical accuracy, the computational burden is incompatible with the rapid iteration cycles often required during network planning and real-time reconfiguration.

Traditional AP deployment strategies, such as empirical path loss models or heuristic optimization methods, face limitations that include oversimplified channel assumptions, heavy measurement requirements, or susceptibility to local optima. Purely data-driven machine learning (ML) approaches offer faster prediction once trained but often struggle to generalize to spatially varying tunnel characteristics and may fail to capture critical physical effects.

To address these challenges, we propose augmenting the limited set of PWE-simulated path loss (PL) maps with a conditional generative adversarial network (cGAN) that learns to generate high-resolution, physically consistent PL distributions for untested AP positions. This PL data augmentation substantially reduces the simulation time while maintaining the physical realism required for CBTC-grade network planning. The augmented dataset enables a deep reinforcement learning (DRL) agent to explore a richer set of AP configurations without incurring the prohibitive cost of exhaustive PWE simulations.

We adopt a dueling deep Q-network (Dueling DQN) as the optimization engine, which decouples state value and action advantage estimation to improve convergence speed and exploration–exploitation balance in complex tunnel environments~\cite{Wang2016Dueling,AlMahamid2021Reinforcement}. In addition to the standard DRL formulation, we define a CBTC-specific interaction environment and discrete action space tailored to AP deployment in tunnels, ensuring that the learned strategies directly address the coverage and latency constraints of railway operations.

\subsection{Prior Work}

In recent years, the CBTC system has been widely used for urban rail transportation~\cite{Jiao2022AG5G}. Access points, which connect wireless devices to the wired network, play a vital role in the CBTC system. Once the train crosses the signal coverage area of the AP, bidirectional communication is established between the train and the ground~\cite{Wen2018Access}. When deploying APs, it is critical to find AP positions that ensure efficient and stable communication, particularly in railway tunnels, which are dark and enclosed spaces. Hence, an algorithm tailored to optimize AP placement in railway tunnel environments is imperative. In wireless communication systems, there are two modules that must be addressed when developing an AP placement optimization algorithm: a channel model and an optimization algorithm~\cite{Zhang2018Physics}.

For the channel model, an extensively utilized method in AP positioning optimization within the wireless domain, called the empirical path loss model, can be adopted~\cite{Sherali1996Optimal,AguadoAgelet2002Optimization}, and it must be simple, accurate and general~\cite{Zhou2024Environment}. The model relies firmly on measurement results~\cite{Myagmardulam2021Path,Popoola2018Comparative}, thus requiring massive measurements. In contrast, a PWE method based on computational simulation can also be accurately applied to channel model development~\cite{Li2024Large,Noori2005New}. This method is suitable for addressing complex terrain and long-distance propagation~\cite{Guan2016Vector}, particularly exhibiting high computational efficiency~\cite{Qin23comparative}. 
However, the PWE method often entails significant computational complexity, particularly when high accuracy is required. In that case, an optimization strategy capable of mitigating computational complexity without significant loss in accuracy thus becomes necessary. Our strategy is to predict explicit and elaborate path loss distribution within our research region by a few PWE-generated path loss values, and several predicting methods have utilized in the process of propagation modeling~\cite{sun2016investigation,qiang2021prediction,saravia2024comparing}. In addition, deep learning-based prediction methods-such as convolutional neural networks (CNNs) and generative adversarial networks (GANs)-have been proven useful~\cite{rafie2022path,marey2022pl}. Meanwhile, GAN, which tends to generate data with random and uncontrollable categories, can be addressed by the conditional generative adversarial network (cGAN), a model that incorporates conditional information to better leverage the strengths of supervised learning in generative tasks, thereby enhancing the matching accuracy between generated data and target classes~\cite{vishwakarma2020comparative,zhang2023rme}. Additionally, cGAN-based prediction approaches have been proven effective and accurate in predicting physical parameters within propagation modeling, such as path loss~\cite{cisse2023irgan}.

Numerous methods exist to optimize the AP positions. A method called Hooke and Jeeves (HJ) optimization is proposed, which moves APs by setting certain step sizes and compares a cost function generated by the path loss~\cite{Zhang2018Physics}. This method is not only effective in simulations, but is also validated in a real subway tunnel environment~\cite{Zhang2018Physics}. In addition, equally distributed placement (EDP), optimal placement (OP) method based on sequential quadratic programming (SQP) to solve the constrained nonlinear optimization problem, hybrid placement (HP) method integrating both the above mentioned approaches, and measurement-based placement (MBP) by placing APs step by step, measuring path loss, and selecting the point with the maximum PL to deploy the next AP are still useful to optimize the placement of the AP along the track~\cite{Saki2020Comprehensive}. Furthermore, the equally distributed placement (EDP) method, particle swarm optimization (PSO) and the combined strategy are proposed to determine the location of APs while deployed in underground mines~\cite{Forooshani2014Optimization}. Meanwhile, the optimization method based on the multi-objective genetic algorithm (MOGA) has also played an optimal role in the placement of wireless local area networks (WLANs)~\cite{Maksuriwong2003Wireless}. In the area of radio wave transmission, plenty of machine leaning (ML) algorithms are used for optimization~\cite{Seretis2022Overview, Huang25Generalizable, Huang25Efficient}. Among them, indoor positioning is achieved using machine learning models, such as learning based on received signal strength (RSS) maps to estimate user locations. This includes both regression methods for calculating user coordinates and classification methods for determining the user's subspace. Furthermore, \cite{Seretis2022Overview, qin2023high} discusses various ML methods applied to radio wave research, which provides inspiration to solve the optimization problem of AP placement in tunnel environments. Among them, a deep Q-learning network (DQN) method belonging to the ML category is applied to the location optimization problem. DQN based on the epsilon-greedy strategy is used to optimize security container placement~\cite{Zhou2023Security}. In addition, DQN is also combined with the whale optimization algorithm (WOA) to address the deployment problem of fiber Bragg grating (FBG) sensors in tunnels~\cite{Liu2024Novel}. Meanwhile, the cloud native network function (CNF) deployment problem is also solved by the DQN algorithm~\cite{Kim2023Deep}. When using DQN for the planning of the path of unmanned aerial vehicles (UAVs)~\cite{Qin2025Physics}, DQN is optimized into the dueling DQN method, achieving better results. Therefore, when solving the AP deployment problem, dueling DQN method can be used as an alternative to the traditional DQN method.

\subsection{Contributions}
Based on the aforementioned facts, we propose a framework based on deep reinforcement learning. We dedicate ourselves to using PWE to generate the electric field distribution in railway tunnel environments and applying DQN network to optimize the AP positions to enhance the received power of the receivers on the trains. To overcome the high computational cost of generating complete PWE datasets, we introduce a conditional GAN-based path loss data augmentation scheme, enabling realistic, high-resolution PL map generation from a small set of simulations. On top of the standard DRL architecture, we design a CBTC-specific interaction environment and action space that reflect operational constraints and latency requirements. The main contributions of this paper are as follows:

\begin{enumerate}
  \item We propose a hybrid AP placement optimization framework that combines high-fidelity path loss data generated by the parabolic wave equation method with a conditional GAN-based data augmentation module. The cGAN learns to generate realistic PL maps for untested AP positions, substantially reducing the computational burden of exhaustive PWE simulations.

  \item We design a CBTC-specific deep reinforcement learning environment, defining tailored state and action spaces that reflect tunnel geometry, coverage requirements, and latency constraints. This customization ensures that the learned AP placement strategies are directly applicable to real-world CBTC operations.

  \item We employ a dueling deep Q-network to perform AP placement optimization, leveraging its decoupled estimation of state value and action advantage to enhance convergence speed, improve exploration–exploitation balance, and avoid suboptimal local solutions.

  \item We conduct comprehensive comparisons against Hooke and Jeeves optimization and traditional DQN methods. Results demonstrate that our cGAN-Dueling DQN framework achieves superior average and worst-case received power performance, validating its ability to produce cost-efficient and robust AP deployment strategies.
\end{enumerate}

\section{Systme Model and Problem Formulation}
\begin{figure}[!h]
  \centering
  \includegraphics[width=3.5in]{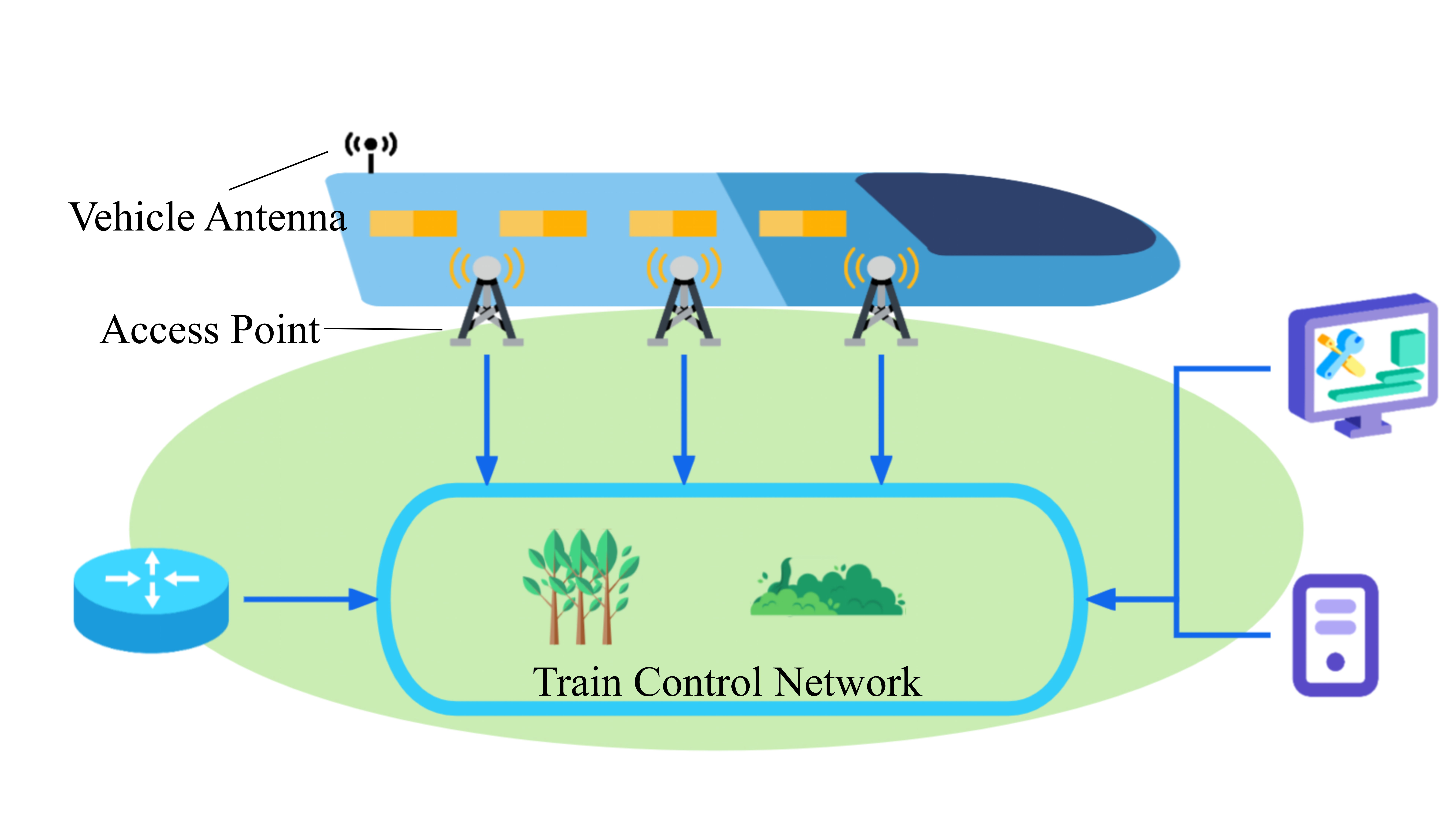}
  \caption{Illustration of communication-based train-control system.}
  \label{fig_1}
  \end{figure}
\begin{figure}[!h]
  \centering
  \includegraphics[width=3.3in]{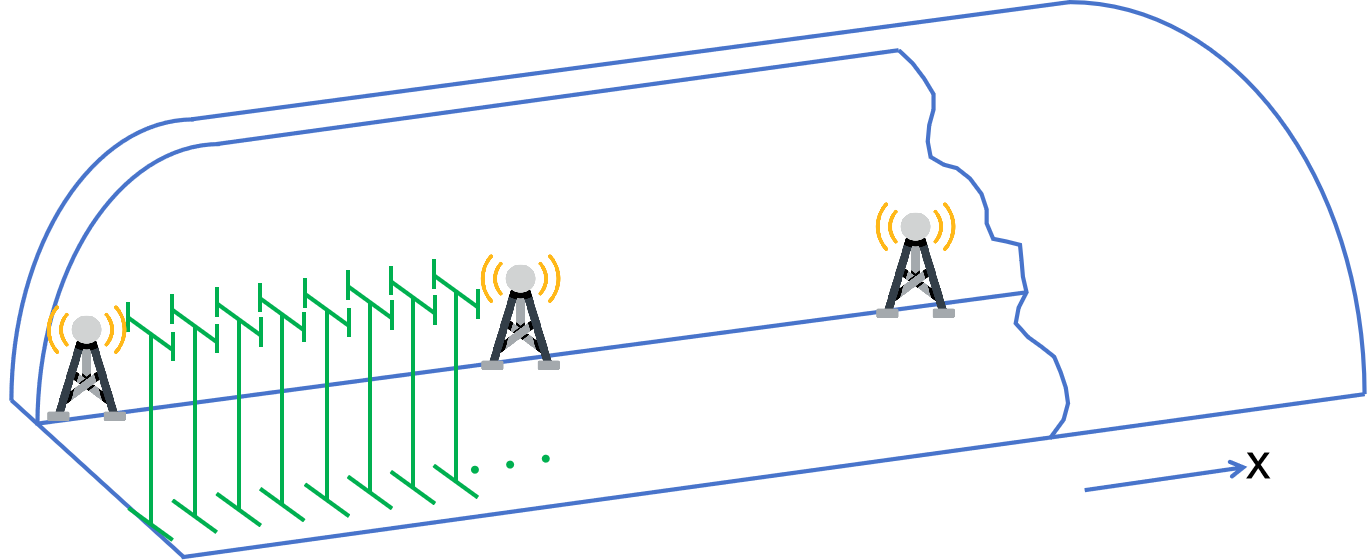}
  \caption{Simplified CBTC tunnel model.}
  \label{fig_2}
  \end{figure}

The key components of a CBTC system include on-board controllers, wayside infrastructure, and communication networks (e.g., LTE, Wi-Fi), as shown in \autoref{fig_1}. Access point is the core equipment of wayside infrastructure to complete real-time interaction with the speeding train. To receive the signal transmitted from the wayside APs, a vehicle antenna is deployed in the specific location on the train. In a train tunnel scenario, a certain number of access points are installed every few hundred meters on either side of the rail tracks, with transmitting antennas attached to them. As demonstrated in \autoref{fig_2}, when a train moves along the rail tracks, the vehicle antenna can be seen as a series of receivers placed along the route of the train since the speed of the train is quite fast. 
The entire network of access points is responsible for serving all the receivers along the train tunnel. During signal transmission, path loss occurs due to signal attenuation, which means that the signal power arrived at the remote receivers could not meet the demand for real-time communication between access points and the speeding train. Therefore, the optimization problem is to deploy the access points in a way that minimizes the average path loss at each receiver points and prevent extremely high path loss somewhere remote of the receiver points from happening.
\subsection{Propagation Environment Model}
In this optimization problem, we consider a relatively simple circumstance with a crooked tunnel and the access points are placed along one-dimensional line. With the tunnel topology and the specifications of transmitting antennas and receiving antennas, the received power was obtained at each receiver point by using the method of PWE. The received power is taken as the power received by the antenna on the train, and for a given location of a access point, the path loss(in dB) at the \(i^{\mathrm{th}}\) receiver point can then be calculated by:
\begin{equation}
  \mathrm{pl}_i = P_0 - P_i
  \end{equation}where \(P_0\) is the transmit power from the access point, \(P_i\) is the power received at the \(i^{\mathrm{th}}\) receiver location.

When there are multiple access points deployed in the tunnel, each AP will serve every receiver points along the curved tunnel, however, the eventually signal behavior of each receiver point is determined by the largest power transmitted from one of the access points, corresponding to lowest path loss. Therefore, for a specific deployment of certain number of access points, the path loss at each receiver point is given by:
\begin{equation}
  \mathrm{PL}_i(\mathbf{X}) = \min_j \{\mathrm{pl}_{i,j}(x_j)\}
  \end{equation}where \(\mathrm{pl}_{i,j}(x_j)\) is the path loss arrived at the \(i^{\mathrm{th}}\) receiver location generated by \(j^{\mathrm{th}}\) AP. \(\mathrm{PL}_i(\mathbf{X})\) represents the ultimately path loss at each receiver point under current AP positioning. \(\mathbf{X}=(x_1,x_2,\dots,x_n)\) denotes the coordinate of \(n\) access points along the tunnel direction.
  \subsection{Optimization Model}

  In order to obtain the optimal spacial position of access points mathematically, cost functions of path loss regarding to the deployment position of APs should be formulated. We construct two type of cost functions that represent the signal power received by receiver points: average path loss and maximum path loss. 

  The first type of cost function is defined as
    \begin{equation}
    g_1(\mathbf{X}) = \frac{1}{m} \sum_{i=1}^{m} \mathrm{PL}_i(\mathbf{X})
    \end{equation}
    where \(m\) is the total number of sampling points of the receiver. This kind of cost funciton tend to appraisal the average behavior of the signal power arrived at each receiver points. It aims to elevate the overall condition of the communication system.
    However, it has a drawback that can not prevent some of the remote receiver points from having extremely high path loss, which is bad for the whole system.

  The second type cost function is defined as
    \begin{equation}
    g_2(\mathbf{X}) = \max_{i} \{\mathrm{PL}_i(\mathbf{X})\}
    \end{equation}
    This kind of cost function ensures that the worst circumstance of the path loss still below some number of threshold, compensating the weakness of the first cost function.

  In this train communication system in tunnel, every position that vehicle antenna passed through should at least remain some value of signal power, that is, the path loss of each receiver point should always smaller than a treshold value, in order to keep the train running safely:

  \begin{equation}
    \mathrm{PL}_i(\mathbf{X}) \leq th_i
    \label{th}
    \end{equation}where \(th_i\) is the maximum tolorated path loss of the \(i^{\mathrm{th}}\) receiver point. This constraint can be incorporated in the previous cost function as a penalty term. If the path loss of \(i^{\mathrm{th}}\) receiver point exceeds the presuppose threshold, penalty should be add to the cost function to make it larger. If the path loss does not break the limit, then nothing would be added. By adopting this method, our model ensures the signal quality of every receiver point and improve the convergence of the optimization process. Eventually, the two cost function can then be represented as:
    \begin{equation}
      f_1(\mathbf{X}) = \frac{1}{m} \sum_{i=1}^{m} \big[ \mathrm{PL}_i(\mathbf{X})  + \lambda_i \max\{0, \mathrm{PL}_i(\mathbf{X}) - th_i \} \big]
      \label{f1}
      \end{equation}
      
      \begin{equation}
      f_2(\mathbf{X}) = \max_{i} \big[ \mathrm{PL}_i(\mathbf{X}) + \lambda_i \max\{0, \mathrm{PL}_i(\mathbf{X}) - th_i \} \big]
      \label{f2}
      \end{equation}where \(\lambda_i\) is the penalty coefficient for violating prescribed path loss threshold at the \(i^{\mathrm{th}}\) receiver point. 

    In this paper,a convex combination of these two cost functions is adopted. We defind \(\alpha\) as the convex combinational weight, and its value can be adjusted by system user at the range of \([0,1]\) according to the operating requirements. Thus, the convex combination of cost functions can be formulated as:
\begin{equation}
\min\  \alpha f_1(\mathbf{X}) + (1 - \alpha) f_2(\mathbf{X}) 
\label{cost}
\end{equation}
subject to \[ 0 \leq x_j \leq L_j,\ j = 1, \dots, n \]
where \(L_j\) denotes the total length of the curved tunnel, providing limitation to this convex combination optimizaiton problem. This combination method endows our system with more comprehensive communication behaviour, guaranteeing all received power overtop some value of threshold while maintaining a  relatively low average path loss simultaneously.

We summarize the optimization problem of AP location as the following formula:

\begin{equation}
		(\text{P1}):   \\
			\min_{\mathbf{X}} [\alpha f_1(\mathbf{X}) + (1 - \alpha) f_2(\mathbf{X})]\\ 
		\label{eq:P1}
	\end{equation}
	
    \begin{align}
		\re{\text{s.t.}} & \re{f_1(\mathbf{X}) = \frac{1}{m} \sum_{i = 1}^{m} [\text{PL}_i(\mathbf{X}) + \lambda_i \max\{0, \text{PL}_i(\mathbf{X}) - \text{th}_i\}] }\tag{\ref{eq:P1}a}	\\
		&\re{f_2(\mathbf{X}) = \max_i [\text{PL}_i(\mathbf{X}) + \lambda_i \max\{0, \text{PL}_i(\mathbf{X}) - \text{th}_i\}]}\tag{\ref{eq:P1}b}	\\
		& \re{\text{PL}_i(\mathbf{X}) = \min_j \{\text{pl}_{i,j}(x_j)\}}\tag{\ref{eq:P1}c}\\
		& \re{\text{pl}_{i,j}(x_j) = P_0 - P_{i,j}}\tag{\ref{eq:P1}d}\\
		& \re{0 \leq x_j \leq L_j, \quad \forall j = 1, 2, \dots, n}\tag{\ref{eq:P1}e}\\
		& \re{\mathrm{PL}_i(\mathbf{X}) \leq th_i} \tag{\ref{eq:P1}f} \\
		& \re{\alpha \in [0, 1]} \tag{\ref{eq:P1}g}
	\end{align}


\section{Deep Q-Learning Network}
\subsection{Basic Conceptions}
Reinforcement learning is a type of machine learning method that involves learning how to map states to actions in order to maximize the rewards obtained. An agent in this context needs to continuously experiment within its environment, using feedback (rewards) from the environment to continuously optimize the mapping between states and actions.

Markov Decision Process (MDP) is a mathematical modeling framework for RL. It is a standard model for describing sequential decision-making problems, defined by the five-tuple $\mathcal{M} = (S, A, P, R, \gamma)$:

\begin{itemize}
    \item \textbf{State (S)}: The set of discrete/continuous states of the environment (e.g., the joint angles of a robot, the game screen).
    \item \textbf{Action (A)}: The set of actions that the agent can perform (e.g., movement commands of a robotic arm, game controller operations).
    \item \textbf{Transition Probability (P)}: The conditional probability of state transition $P(s' | s, a)$ (the dynamic model of the environment).
    \item \textbf{Reward (R)}: The immediate reward function $R(s, a, s')$ (defining the task objective).
    \item \textbf{Discount Factor ($\gamma$)}: The discount factor for future rewards (balancing short-term and long-term benefits).
\end{itemize}

Q-learning is a classic model-free algorithm in Reinforcement Learning (RL), used to find the optimal policy in discrete state and action spaces, addressing the Markov decision problems in reinforcement learning. The core of Q-learning is the Q-table (state-action value table), which stores the Q-value \( Q(s, a) \) corresponding to each state \( s \) and action \( a \). The Q-value represents the maximum cumulative reward (discounted) that can be obtained by following the current policy after performing action \( a \) in state \( s \).

The update of Q-values is based on the Bellman Equation:
\begin{equation}
  \begin{split}
  Q(s_t, a_t) &\leftarrow Q(s_t, a_t) \\
  &+ \alpha \left[ r_t + \gamma \max_{a'} Q(s_{t+1}, a') - Q(s_t, a_t) \right]
  \end{split}
  \end{equation} where \( \alpha \) is the learning rate, controls the step size of the update. \( \gamma \) denotes discount factor, balances the importance of current and future rewards (\( 0 \leq \gamma < 1 \)). \( r_t \) is the immediate reward obtained after performing action \( a_t \).

However, the Q-learning algorithm is not suitable for our problem scenario due to its inability to handle high-dimensional state spaces, which may encounter the curse of dimensionality—specifically, the number of required state-action pairs grows exponentially with the increase in the dimensionality of the state space. Therefore, we introduce Deep Q-learning Network, a combination algorithm of Q-learning and artificial neural network. DQN utilizes neural network to replace the Q-table in Q-learning algortithm with the network update mode:
\begin{equation}
  \begin{split}
  Q(S_t, A_t, w) &\leftarrow Q(S_t, A_t, w) + \\ &\alpha \left[ R_{t+1} + \gamma \max_{a} \hat{q}(s_{t+1}, a_t, w) - Q(S_t, A_t, w) \right]
  \end{split}
  \end{equation} where \(w\) is the network parameter.

  The \(\varepsilon\)-greedy strategy commonly used in DQN is a key mechanism for balancing exploration and exploitation. This strategy selects actions randomly with probability \(\varepsilon\) to explore actions that are not well understood, thus preventing the agent from falling into local optima; at the same time, it exploits with probability \(1 - \varepsilon\) by choosing the action with the highest estimated value from the current Q-network. The mathematical expression is as follows:

\begin{equation}
\pi(a|s) = 
\begin{cases} 
\arg\max_{a'} Q(s, a'; \theta) & \text{with } 1 - \varepsilon \\
\text{random action} & \text{with } \varepsilon 
\end{cases}
\end{equation}When executing, for the current state \(s\), a random number \(\xi\) is generated. If \(\xi < \varepsilon\), an action is selected randomly from the action space; if \(\xi \geq \varepsilon\), the action that maximizes \(Q(s, a; \theta)\) is chosen. In our work, we adopt exponential decay method to update \(\varepsilon\):
\begin{equation}
  \varepsilon_t = \varepsilon_{\min} + \left( \varepsilon_{\max} - \varepsilon_{\min} \right) \cdot e^{-\lambda t}
  \end{equation}This strategy is significant as it not only prevents the agent from relying too early on suboptimal strategies but also ensures convergence through exploration.

\subsection{CBTC Environment Definition for Deep Q-Network}
In our context of the railway tunnel communication system, the AP position problem is a Markov decision problem, for the next state AP position is solely dependent on the current state AP position and its consequent path loss generated at each receiver point. Here in our problem setting, a 1500-meters-long train tunnel is considered, and 3 access points are then managed to be placed in this tunnel. At the beginning, we first determine the initial points of positioning, say \([0,500,1000]\), and then define the elemantary process of MDP.

In the current state \(t\), the deployment of three APs with a 3-dimensional positioning vector \(\bm{\langle x_1, x_2, x_3 \rangle}\) produces a cost function \(g\) that we previously defined in Section \text{II}. A state at time 
\(t\) is represented as:
\begin{equation}
  s_t = \begin{pmatrix}
  x_1^{(t)} \\
  x_2^{(t)} \\
  x_3^{(t)} \\
  g^{(t)}
  \end{pmatrix}, \quad 
  \begin{cases} 
  x_i^{(t)} \in [0, L]\\
  g^{(t)} = f(x_1^{(t)}, x_2^{(t)}, x_3^{(t)}) 
  \end{cases}
  \end{equation}where \(x_i^{(t)}\) is the position of the \(i\)-th AP.

  At each state transition step, each AP has three possible movement strategies. In time $t$, the precise positions of the three APs generate a cost function $g^{(t)}$. A deep neural network drives the transition from the current state $t$ to the next state $t+1$ according to the state transition function. Each AP can move left by one step, move right by one step, or remain stationary, corresponding to the following mappings:
\[
\text{Left move} \mapsto -1, \quad \text{Right move} \mapsto +1, \quad \text{No move} \mapsto 0.
\]
Consequently, each state transition involves \textbf{27 possible actions}, represented by the action set:
\begin{equation}
  \begin{aligned}
  \mathcal{A} &= \{-1, 0, +1\}^3 =\\
  & \left\{ (a_1, a_2, a_3) \,\middle|\, a_i \in \{-1, 0, +1\}, \, i = 1, 2, 3 \right\}.
  \end{aligned}
  \end{equation}To ensure compatibility with discrete action reinforcement learning algorithms, the 3D action vector is bijectively mapped to an integer index. This establishes a one-to-one correspondence between action vectors and indices, enabling seamless integration with policy networks. Each action is represented as a 3D vector $\mathbf{a} = (a_1, a_2, a_3)$, where $a_i \in \{-1, 0, +1\}$ denotes the direction of movement of the $i$ th AP. The encoding function $\phi : \{-1, 0, +1\}^3 \rightarrow \{0, 1, \ldots, 26\}$ converts the vector into an integer index $a_{\text{idx}}$ using:
  \begin{equation}
  \phi(a_1, a_2, a_3) = (a_1 + 1) \cdot 9 + (a_2 + 1) \cdot 3 + (a_3 + 1).
  \end{equation}This transformation shifts each component from $\{-1, 0, +1\}$ to $\{0, 1, 2\}$, interprets them as digits in a base-3 numeral system, and computes their base-10 equivalent. The inverse mapping $\phi^{-1} : \{0, 1, \ldots, 26\} \rightarrow \{-1, 0, +1\}^3$ reconstructs the original action vector from the integer index via:
  \begin{align}
  \begin{cases}
  a_1 = \left\lfloor \frac{a_{\text{idx}}}{9} \right\rfloor - 1, \\
  a_2 = \left\lfloor \frac{a_{\text{idx}} \mod 9}{3} \right\rfloor - 1, \\
  a_3 = (a_{\text{idx}}\mod 3) - 1,
  \end{cases}
  \end{align}
  where $\left\lfloor \cdot \right\rfloor$ denotes the floor operation. This process reverses the base-3 decomposition to recover the original action components. The bijective nature of the mapping guarantees invertibility without information loss, while its reliance on arithmetic operations ensures computational efficiency. The integer indices align directly with the discrete action outputs of DQN policies, bridging the gap between multidimensional movement logic and reinforcement learning requirements.

  The state transition function governs how the positions of the three Access Points (APs) evolve from time step $t$ to $t + 1$. For each AP $i$ ($i = 1, 2, 3$), its updated position $x_i^{(t+1)}$ is computed as:
\begin{equation}
x_i^{(t+1)} = \text{clip}\left(x_i^{(t)} + a_i \cdot \Delta, 0, L\right),
\end{equation}
where:
\begin{itemize}
    \item $x_i^{(t)}$ denotes the position of the $i$-th AP at time $t$,
    \item $a_i \in \{-1, 0, +1\}$ represents the discrete movement action for the $i$-th AP (left, stationary, or right),
    \item $\Delta$ is the predefined step size (hyperparameter in neural network),
    \item $L$ is the tunnel length (e.g., 1500 meters),
    \item $\text{clip}(v, v_{\min}, v_{\max})$ enforces boundary constraints by clipping the value $v$ to the range $[v_{\min}, v_{\max}]$:
    \begin{equation}
    \text{clip}(v, v_{\min}, v_{\max}) \triangleq \max(v_{\min}, \min(v, v_{\max})).
    \end{equation}
\end{itemize}This function ensures that AP movements remain confined within the spatial limits of the tunnel ($0 \leq x_i^{(t+1)} \leq L$). The term $a_i \cdot \Delta$ adjusts the position by one step left ($-\Delta$), right ($+\Delta$), or no movement ($0$), depending on the action. The clipping operation guarantees physical feasibility, preventing APs from moving beyond the tunnel boundaries.

The immediate reward \( r_t \) is defined as the negative value of the cost function at the next state:
\begin{equation}
r_t = -g^{(t+1)}
\end{equation}
where \( g^{(t+1)} \) quantifies the coverage quality after transitioning to state \( t + 1 \). This incentivizes the agent to minimize the cost function, as higher rewards correspond to lower coverage costs.

An episode terminates when the coverage quality meets the predefined threshold:
\begin{equation}
\text{Terminate if } g^{(t)} < 20
\label{terminate}
\end{equation}
This ensures the process stops once the AP placements achieve satisfactory coverage (cost below 20), concluding the optimization episode.
 \begin{table}[htbp]
    \centering
    \small
    \begin{tabular}{cc}
        \toprule 
        Parameter & Value \\
        \midrule 
        Learning rate & 0.001 \\
        $\gamma$ & 0.995 \\
        $\varepsilon$ (epsilon) & 1 \\
        $\varepsilon$-decay (epsilon decay) & 0.995 \\
        $\varepsilon$-min (minimal epsilon) & 0.01 \\
        State size & 4 \\
        Action size & 27 \\
        Episodes & 1500 \\
        Batch size & 64 \\
        \bottomrule 
    \end{tabular}
    \caption{DRL algorithm parameters.} 
    \label{tab:DRL_parameters} 
\end{table}
Thus far, the AP placement optimization model for train tunnels and its DQN optimization algorithm have been fully presented. The simplified illustrative expressions are as follows:

\[
\begin{pmatrix}
x_1^{(t)} \\
x_2^{(t)} \\
x_3^{(t)} \\
g^{(t)}
\end{pmatrix}
\xrightarrow{\phi^{-1}(a_{\text{idx}})}  
\begin{pmatrix}
x_1^{(t+1)} \\
x_2^{(t+1)} \\
x_3^{(t+1)} \\
g^{(t+1)}
\end{pmatrix}
\]

\begin{figure}[!h]
	\centering
	\includegraphics[width=3.2in]{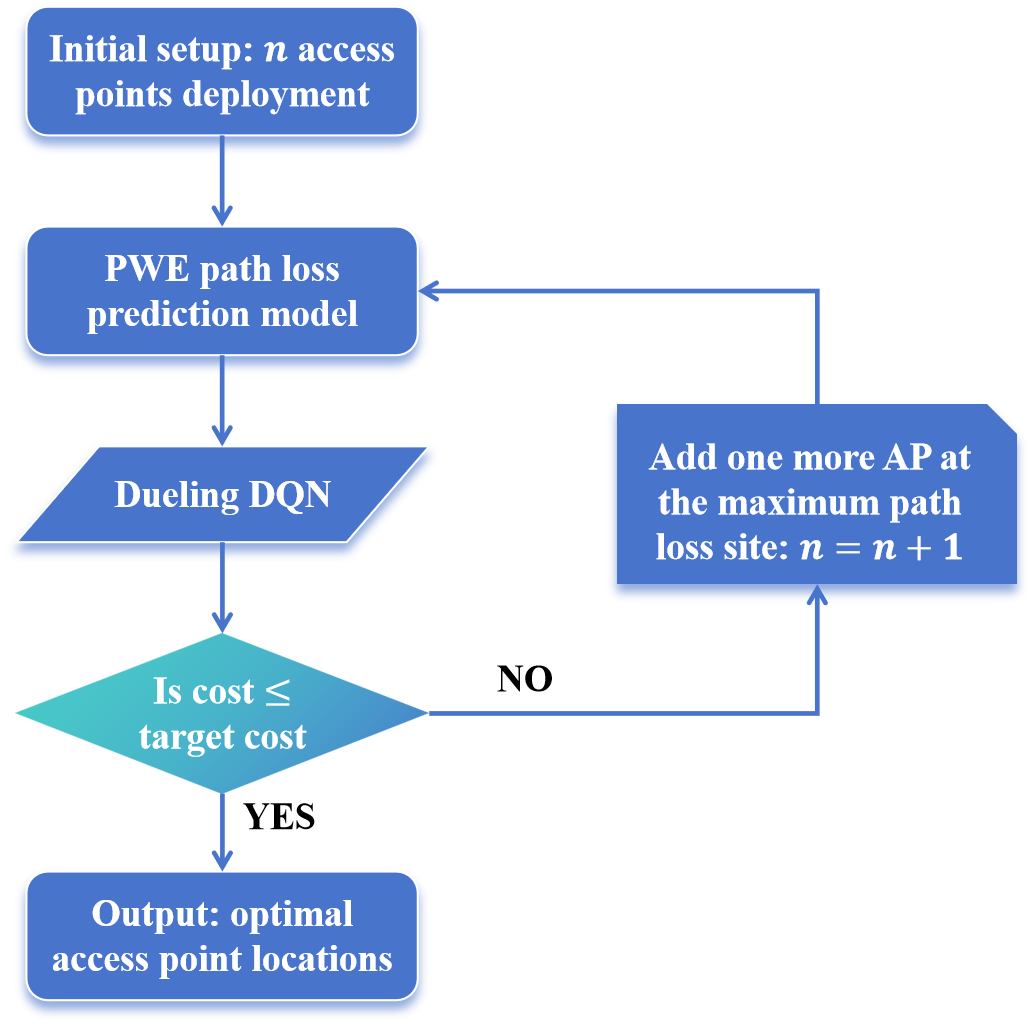}
	\caption{Diagram of the optimization procedure.}
	\label{fig_procedure}
\end{figure}

\medskip  
\noindent  
with transitions governed by \( x_i^{(t+1)} = \text{clip}\left(x_i^{(t)} + a_i \Delta\right) \). The general procedure for finding optimal access point locations is shown in \autoref{fig_procedure}. In our problem setting, we fix 3 access points for convenience along a 1500m-long curved train tunnel.

\begin{figure*}[!h]
	\centering
	\includegraphics[width=6.5in]{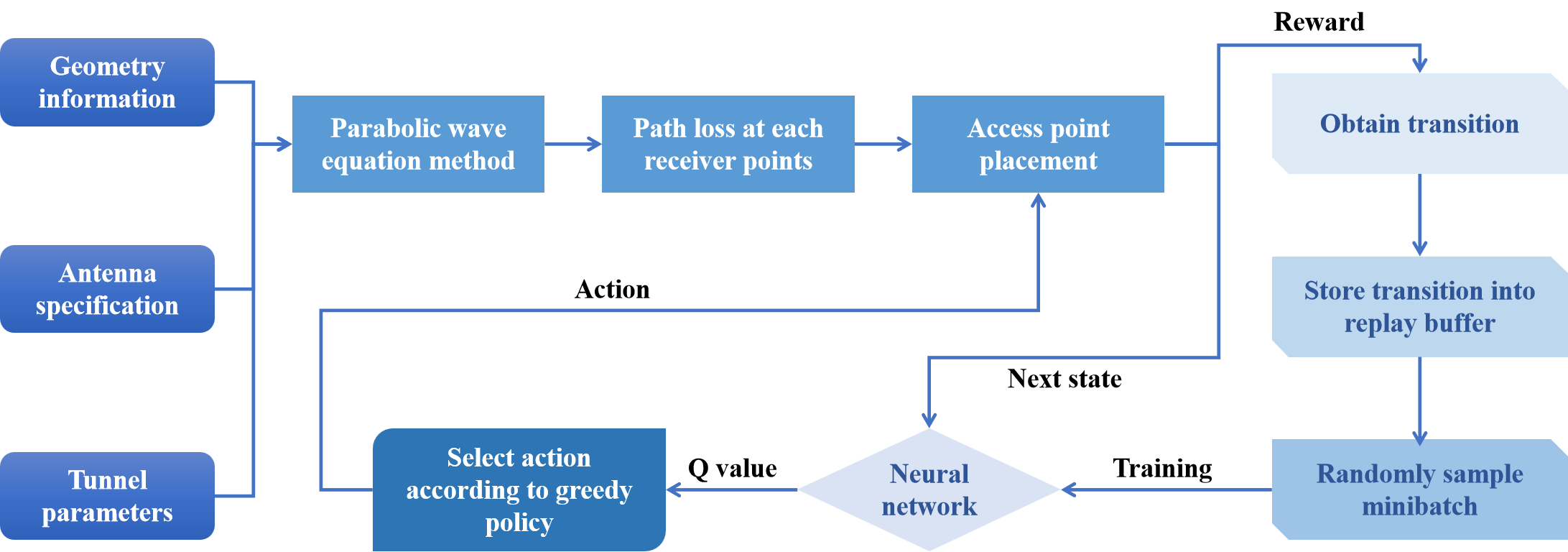}
	\caption{Framework of the proposed approach of AP placement for CBTC systems.}
	\label{fig_framework}
\end{figure*}

\section{Proposed AP Placement Optimization Approach for CBTC Systems}

In our tunnel AP deployment optimization problem, the standard DQN faces two core challenges: First, the traditional Q-network couples the inherent state value with relative action advantages, making it difficult to accurately distinguish subtle differences between actions, especially during fine-tuning stages when AP positions approach optimality. When the spacing between three APs is close to the optimal configuration, the differences in the improvement of the cost function \(g\) among 27 possible actions may be minor. Under such conditions, standard DQN exhibits oscillatory Q-DRL algorithms. Second, in sparse reward environments (where reward signals flatten as \(g\) approaches optimal values), traditional architectures struggle to capture the intrinsic value of states, leading to ambiguous policy updates. Based on the potential risks of the standard DQN mentioned above, we propose an improved method, Dueling DQN, to improve our neural network. The framework of the proposed approach of AP placement strategy is shown in \autoref{fig_framework}.

\begin{figure}[htbp]
	\centering
	\includegraphics[width=3.3in]{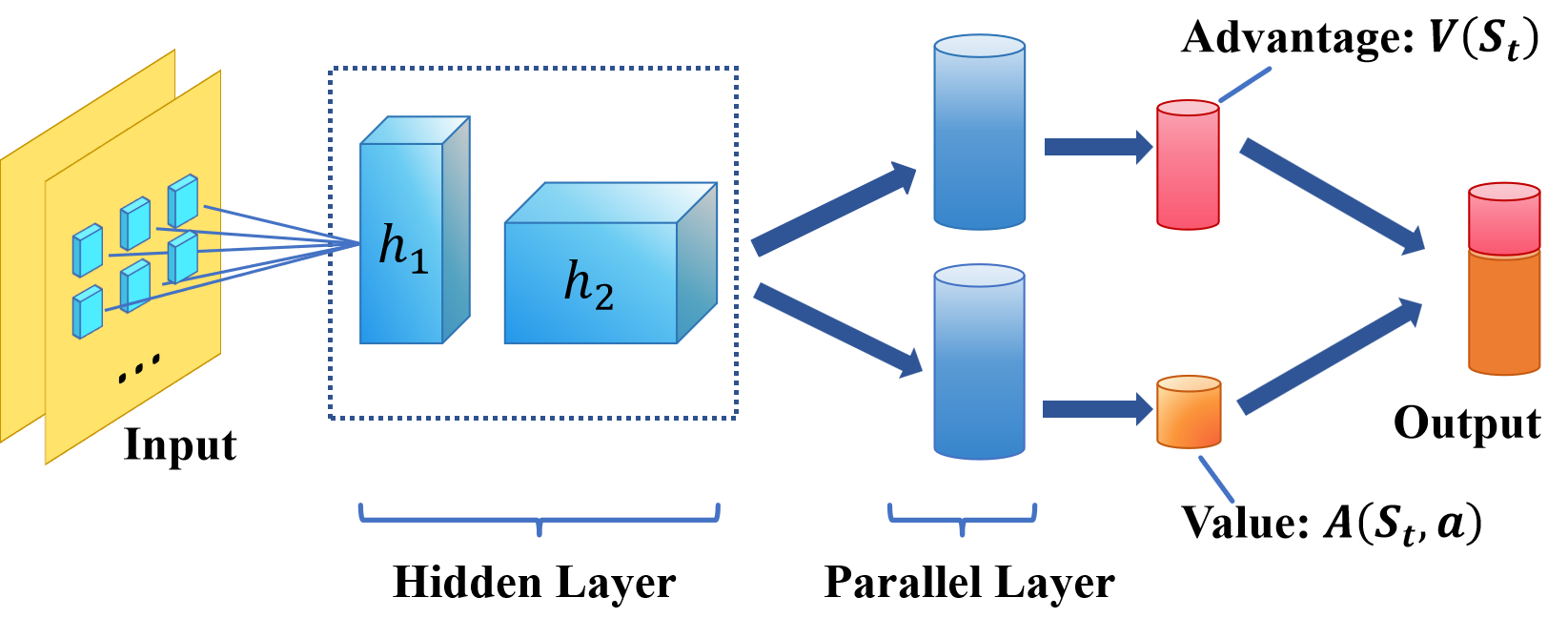}
	\caption{Diagram of the dueling DQN for the AP optimization problem.}
	\label{fig_DDQN}
\end{figure}

\begin{table*}[htbp]
    \centering
    \small 
    \setlength{\tabcolsep}{8pt}
    \begin{tabular}{ccccc}
        \toprule
        \makecell{Optimization \\ Methods} & 
        \makecell{Cost Function Value \\ for Initial AP Positions} & 
        \makecell{Optimized Cost \\ Function Values} & 
        \makecell{Percentage \\ Improvement} & 
        \makecell{Final AP \\ Positions} \\
        \midrule
        HJ Optimization & \multirow{3}{*}{276.83} & 90.02 & 67.5\% & \(\textbf{X}=[0, 483, 979]\) \\ 
        Traditional DQN &  & 83.27 & 69.9\% & \(\textbf{X} = [0, 238, 714]\) \\ 
        Dueling DQN &  & 54.24 & 80.4\% & \(\textbf{X} = [0, 155, 658]\) \\
        \bottomrule
    \end{tabular}
    \caption{Optimization results comparison.}
    \label{tab:optimization_results}
    \vspace{3pt}
    
    \small
    \setlength{\tabcolsep}{8pt}
    \begin{tabular}{ccccc}
        \toprule
        \makecell{Weight Parameter \\ $\alpha$} & 
        \makecell{Optimized AP \\ Positions} & 
        \makecell{Optimized Cost \\ Function Value} & 
        \makecell{Unoptimized Cost \\ Function Value} & 
        \makecell{Percentage \\ Improvement} \\
        \midrule
        0 & \(\textbf{X}=[0, 130, 626]\) & 80.23 & 526.51 & 84.8\% \\
        0.5 & \(\textbf{X}=[0, 155, 658]\) & 54.24 & 276.83 & 80.4\% \\
        1 & \(\textbf{X}=[1, 192, 713]\) & 19.70 & 27.15 & 27.4\% \\
        \bottomrule
    \end{tabular}
    \caption{Parameter adjustment comparison.} 
    \label{tab:parameter_adjust}
\end{table*}

As illustrated in \autoref{fig_DDQN}, the innovation of Dueling DQN lies in decoupling the Q-value function into a linear combination of a state value function \( V(s) \) and an action advantage function \( A(s, a) \):
\begin{equation} Q(s, a) = V(s) + \left( A(s, a) - \frac{1}{|\mathcal{A}|} \sum_{a'} A(s, a') \right) 
\label{ddqneq}
\end{equation}
This decomposition carries profound physical implications: \( V(s) \) characterizes the inherent communication potential of the current AP layout in the tunnel, while \( A(s, a) \) reflects the relative improvement from specific movement commands. For example, when three APs are nearly equidistant (where \( V(s) \) peaks), any action that disrupts this balance generates negative advantages in \( A(s, a) \), even if such actions temporarily enhance local signal strength. This decoupling enables for clearer identification of critical state features.

The network architecture of dueling DQN achieves refined modeling of AP deployment strategies by decoupling state value estimation and action advantage learning. As shown in \autoref{fig_DDQN}, the network consists of a shared feature extraction layer, parallel value advantage evaluation layers, and a normalized output layer. Its design strictly follows the progressive principle of "feature sharing, value decoupling, and advantage normalization." The input layer receives a 4-dimensional state vector \( s_t = [x_1, x_2, x_3, g_t] \), where \( x_i \in [0, 1500] \) represents the AP location coordinates, and \( g_t \) is the current cost function. The forward propagation process begins with nonlinear feature abstraction through two fully connected layers:
\[
\begin{aligned}
h_1 &= \text{ReLU}(W_1 s_t + b_1) \\
h_2 &= \text{ReLU}(W_2 h_1 + b_2)
\end{aligned}
\]
Here, \( W_1 \in \mathbb{R}^{64 \times 4} \) and \( W_2 \in \mathbb{R}^{64 \times 64} \) are shared weight matrices. The ReLU activation function ensures that the network captures non-linear spatial relationships between AP locations, such as the compound effects between adjacent APs on signal attenuation. The 64-dimensional vector \( h_2 \) from the shared feature layer is then decoupled into two parallel branches: the value stream \( V(s_t) \), which evaluates the inherent quality of the current AP layout, and the advantage stream \( A(s_t, a) \), which quantifies the improvement potential of each action relative to the average. This decoupling mechanism allows the network to independently learn two critical types of information—the value stream focuses on identifying high-quality layout patterns aligned with communication engineering principles (e.g., equidistant triangular distributions), while the advantage stream refines the subtle utility differences among 27 possible actions.

The output layer generates the final Q-value through an aggregation formula demonstrated in \autoref{ddqneq}. The core innovation lies in the normalization of the advantage term. By subtracting the mean of action advantages \( \frac{1}{|\mathcal{A}|} \sum_{a'} A(s_t, a') \), the network enforces the constraint:
\begin{equation}
\sum_{a'} \left[ Q(s_t, a) - V(s_t) \right] = 0
\end{equation}
This ensures strict alignment between the Q-value baseline and the inherent state value. In problem practice, this constraint effectively suppresses the oscillation of Q values observed in traditional DQN during the fine-tuning of actions: When the AP positions approach optimality, the advantage values of actions converge and the Q values are governed primarily by \( V(s_t) \). This prevents policy updates from deviating from stable regions due to noise perturbations.

\begin{figure*}[t]
  \centering
  \includegraphics[width=\textwidth]{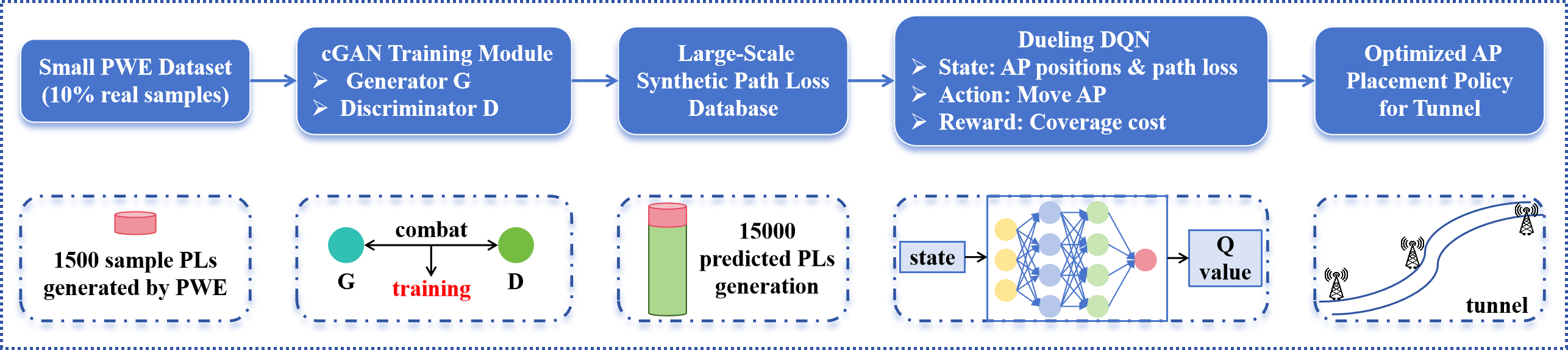}
  \caption{Overall pipeline of cGAN-augmented DRL for AP placement: (i) limited PWE simulations provide high-fidelity PL maps, (ii) cGAN learns to map coarse PL inputs (and AP mask) to fine-resolution PL, (iii) the generated database fuels a Dueling DQN agent for placement optimization.}
  \label{fig:cgandrl_flowchart}
\end{figure*}

\begin{figure*}[t]
    \centering
    \subfloat[]{%
        \includegraphics[width=0.32\textwidth]{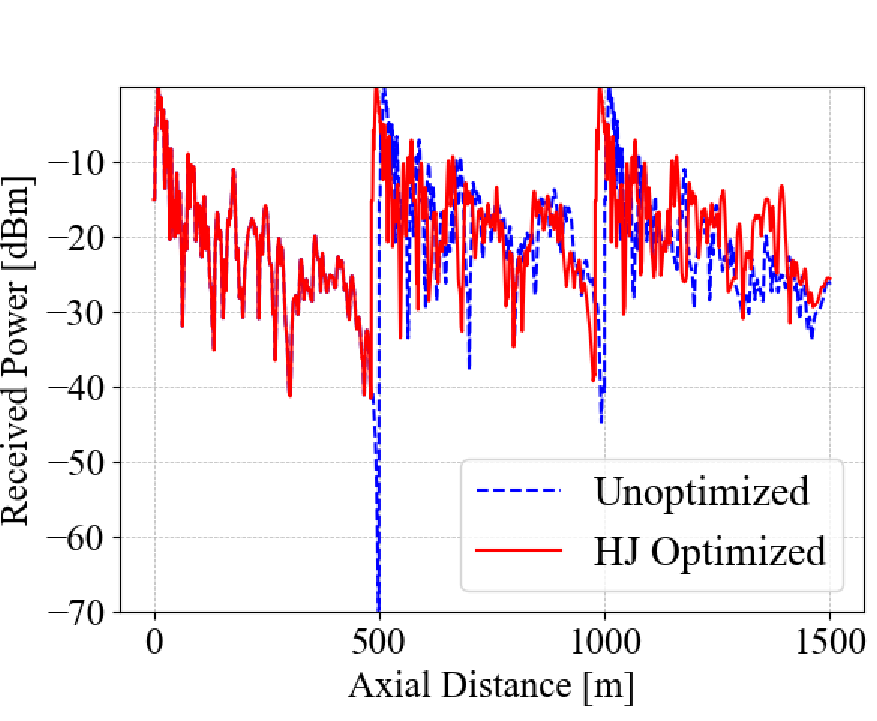}
    }\hspace{-1mm}
    \subfloat[]{%
        \includegraphics[width=0.32\textwidth]{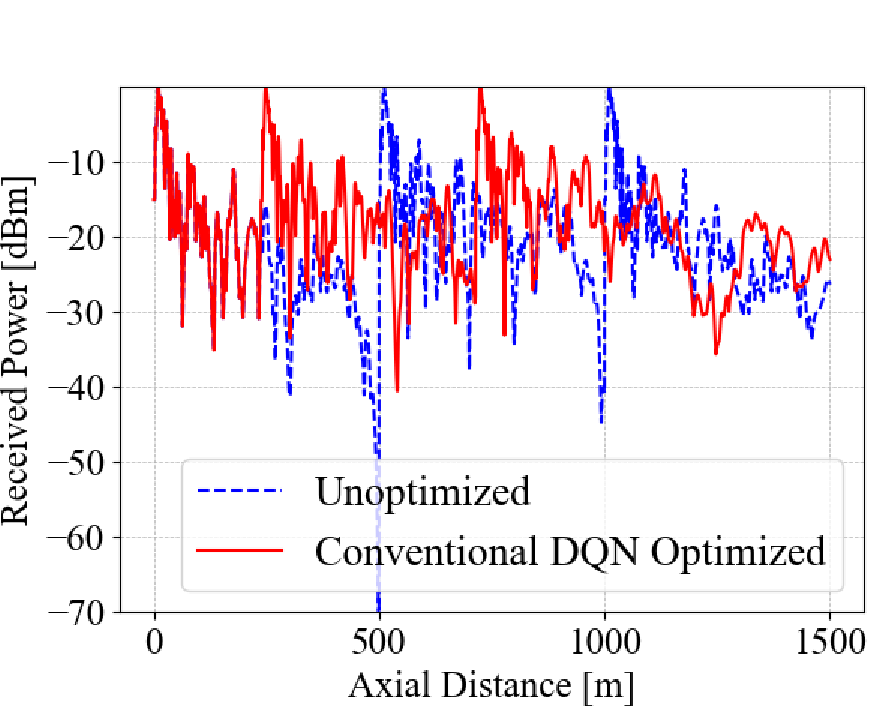}
    }\hspace{-1mm}
    \subfloat[]{%
        \includegraphics[width=0.32\textwidth]{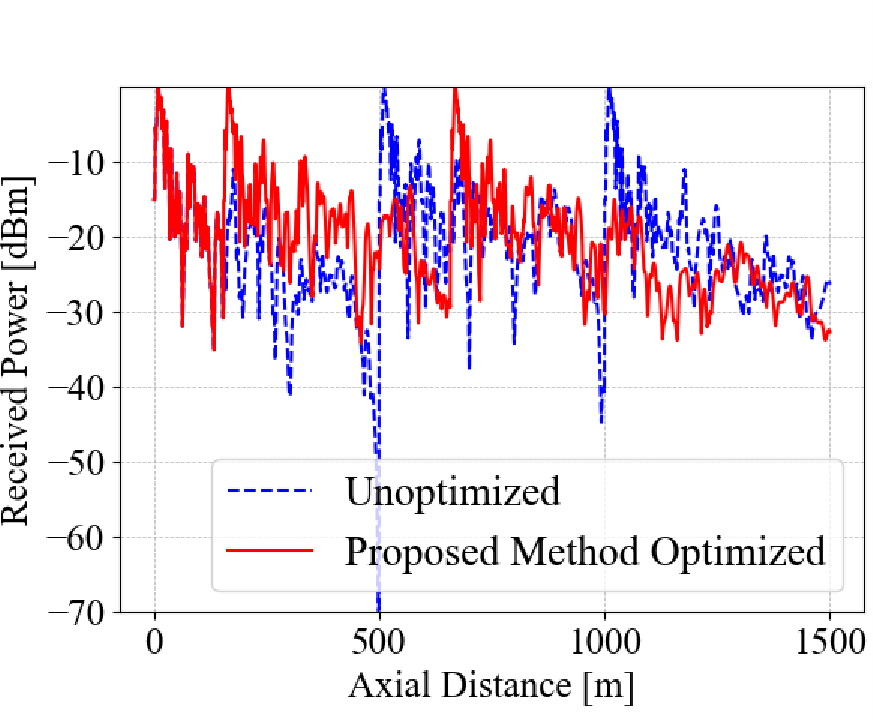}
    }
    \caption{Optimization results of three different methods for AP placement in tunnels: 
    (a) Hooke and Jeeves (HJ), (b) DQN, and (c) Dueling DQN.}
    \label{fig:optimization_results}
\end{figure*}

Dueling DQN enhances AP placement optimization by decoupling the Q-value into a state value \( V(s) \) and action advantages \( A(s, a) \). This separation allows \( V(s) \) to holistically evaluate the intrinsic quality of AP configurations (e.g., uniform spacing), while \( A(s, a) \) isolates the relative impact of individual movements. Crucially, normalizing advantages by subtracting their mean (\( A(s, a) - \mathbb{E}[A(s, a)] \)) stabilizes training by anchoring Q-values to \( V(s) \), suppressing noise during fine-tuning phases. In sparse-reward scenarios (for example, near-optimal \( g \)), \( V(s) \) maintains gradient signals to guide exploration, while \( A(s, a) \) amplifies subtle action differences. The architecture thus achieves superior policy precision and faster convergence compared to standard DQN, particularly in identifying spatially optimal AP layouts under complex signal propagation constraints. Algorithm \autoref{alg:DuelingDQN_AP} provides detailed procedure of our proposed method for AP placement.

\section{cGAN-based Path Loss Data Augmentation}

To alleviate the prohibitive computational cost of running parabolic wave equation (PWE) simulations for every possible access point (AP) position, we introduce a conditional Generative Adversarial Network (cGAN) as a data augmentation module. The cGAN is trained on a limited set of high-fidelity PWE-generated path loss (PL) maps and learns to synthesize realistic PL distributions for unseen AP positions. The overall workflow of the cGAN-augmented optimization framework is illustrated in \autoref{fig:cgandrl_flowchart}.

\subsection{Principle of cGAN}

The cGAN consists of a generator $G$ and a discriminator $D$, both conditioned on auxiliary information describing the AP position and tunnel geometry. Let $\mathbf{x} \in \mathbb{R}^{n_x}$ denote a coarse path loss (PL) profile obtained from sparse PWE simulations or low-resolution field measurements, and let $\mathbf{c} \in \mathbb{R}^{n_c}$ represent a condition vector encoding the AP position, tunnel geometry parameters, and other scenario-specific metadata. The generator learns a conditional mapping from $(\mathbf{x}, \mathbf{c})$ to a high-resolution PL prediction $\hat{\mathbf{y}} \in \mathbb{R}^{n_y}$:  
\begin{equation}
    G: (\mathbf{x}, \mathbf{c}) \longrightarrow \hat{\mathbf{y}},
    \label{eq:generator_mapping}
\end{equation}
where $\hat{\mathbf{y}}$ denotes the generated full-resolution PL map covering the valid forward region of the tunnel.

The discriminator $D$ receives either the real PWE-generated PL map $\mathbf{y}$ or the generated output $\hat{\mathbf{y}}$, along with $(\mathbf{x}, \mathbf{c})$, and estimates the probability of the map being real. The adversarial training objective is:
\begin{equation}
\begin{aligned}
\mathcal{L}_{\mathrm{cGAN}}(G,D) 
&= \mathbb{E}_{\mathbf{x},\mathbf{y},\mathbf{c}}
    \big[\log D(\mathbf{x}, \mathbf{y}, \mathbf{c})\big] \\
&\quad + \mathbb{E}_{\mathbf{x},\mathbf{c}}
    \big[\log \big(1 - D(\mathbf{x}, G(\mathbf{x}, \mathbf{c}), \mathbf{c})\big)\big] \\
&\quad + \lambda \, \mathcal{L}_{\mathrm{rec}}(G),
\end{aligned}
\label{eq:cgan_loss}
\end{equation}
where $\mathcal{L}_{\mathrm{rec}}(G)$ is the reconstruction loss:
\begin{equation}
\mathcal{L}_{\mathrm{rec}}(G) = \mathbb{E}_{\mathbf{x},\mathbf{y},\mathbf{c}}\left[\|\mathbf{y} - G(\mathbf{x}, \mathbf{c})\|_1\right],
\label{eq:l1_loss}
\end{equation}
and $\lambda$ balances perceptual realism and numerical accuracy.

In our implementation, $G$ adopts an encoder-decoder architecture with residual connections to capture both large-scale attenuation patterns and fine-grained fading details, while $D$ employs a PatchGAN structure that enforces local realism. Conditioning is injected at multiple layers to ensure physical consistency with the tunnel environment and AP placement.

\subsection{Role in AP Placement Optimization}

Within the AP placement optimization framework, the cGAN serves as a physics-aware surrogate model for the PWE simulator. Once trained, it can generate high-resolution PL maps for any candidate AP position almost instantly, eliminating the need to re-run costly PWE simulations. This capability significantly expands the pool of available PL data, enabling the optimizer to evaluate a much larger number of placement configurations under the same computational budget.

By learning the statistical and physical characteristics of PWE-generated data, the cGAN preserves key propagation features such as waveguide effects, path-dependent attenuation, and localized fading not captured by empirical models. Consequently, the optimizer can make placement decisions based on realistic coverage predictions, while drastically reducing simulation time. This integration effectively bridges the gap between computational electromagnetics and large-scale optimization, providing a scalable and accurate approach to AP deployment in tunnel-based CBTC systems.

\section{Results and Performance Evaluation}

\begin{figure}[htbp]
	\centering
	\includegraphics[width=3.5in]{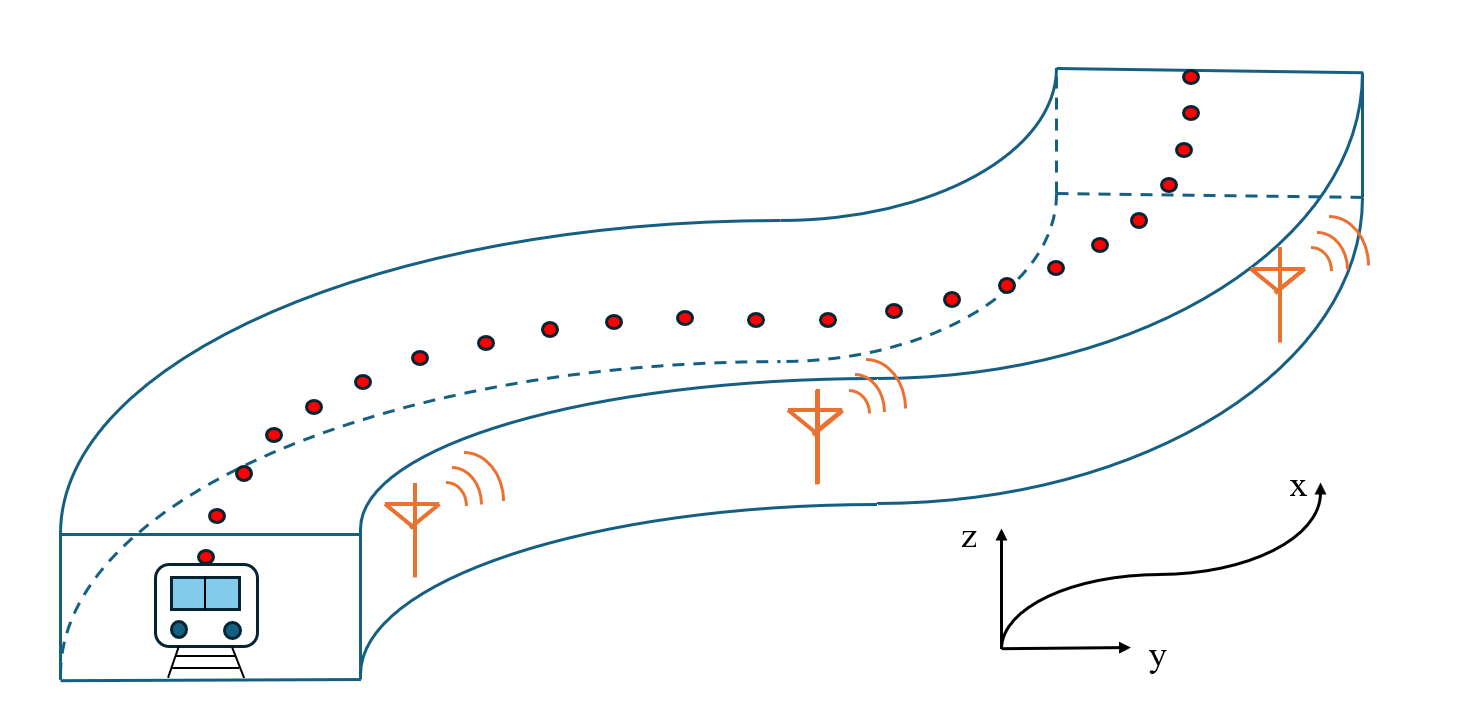}
	\caption{The geometry of the test tunnel.}
	\label{fig_geo}
\end{figure}

In this section, we evaluate the effectiveness of our proposed approach by optimizing the positions of three APs in a 1500m railway tunnel. \autoref{fig_geo} illustrates the geometry of the test tunnel. The tunnel curves right in the first half and left in the second half. The radius of curvature for both is 477.5 m. The electrical parameters of the tunnel walls are \(\epsilon_r=5\) and \(\sigma_0=0.01S/m\).

\begin{figure}[htbp] 
    \centering 
    \includegraphics[width=0.5\textwidth]{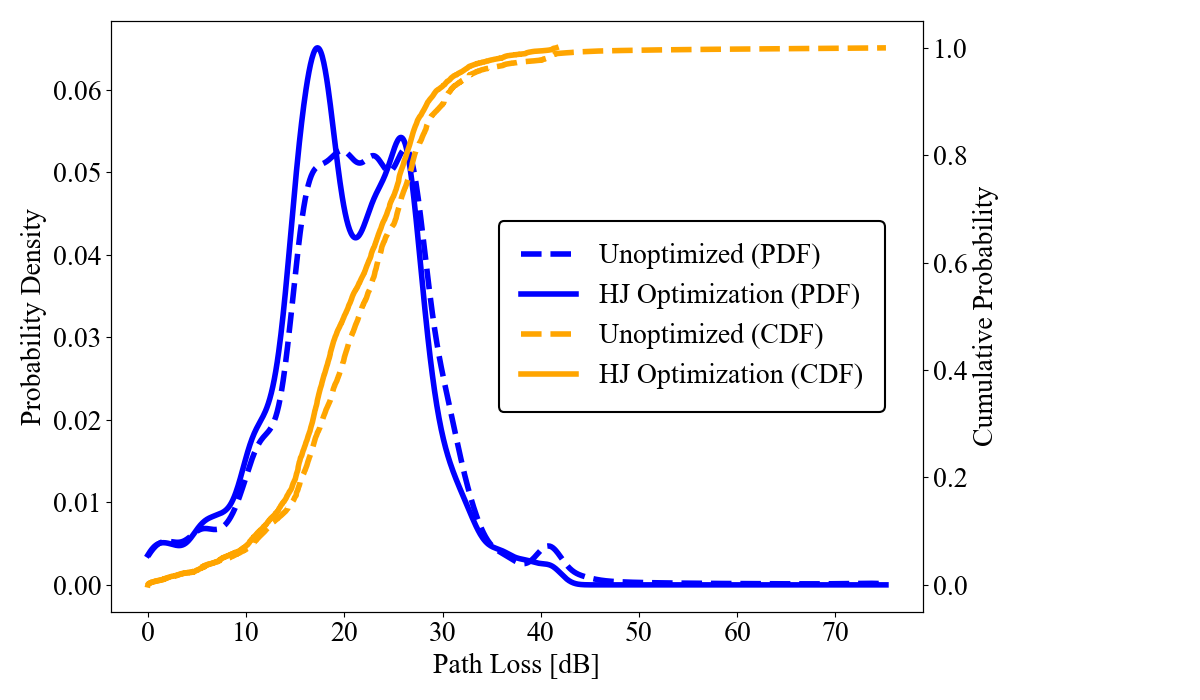} 
    \caption{PDF and CDF of HJ method.} 
    \label{figgl1} 
\end{figure}

\begin{figure}[htbp] 
    \centering 
    \includegraphics[width=0.5\textwidth]{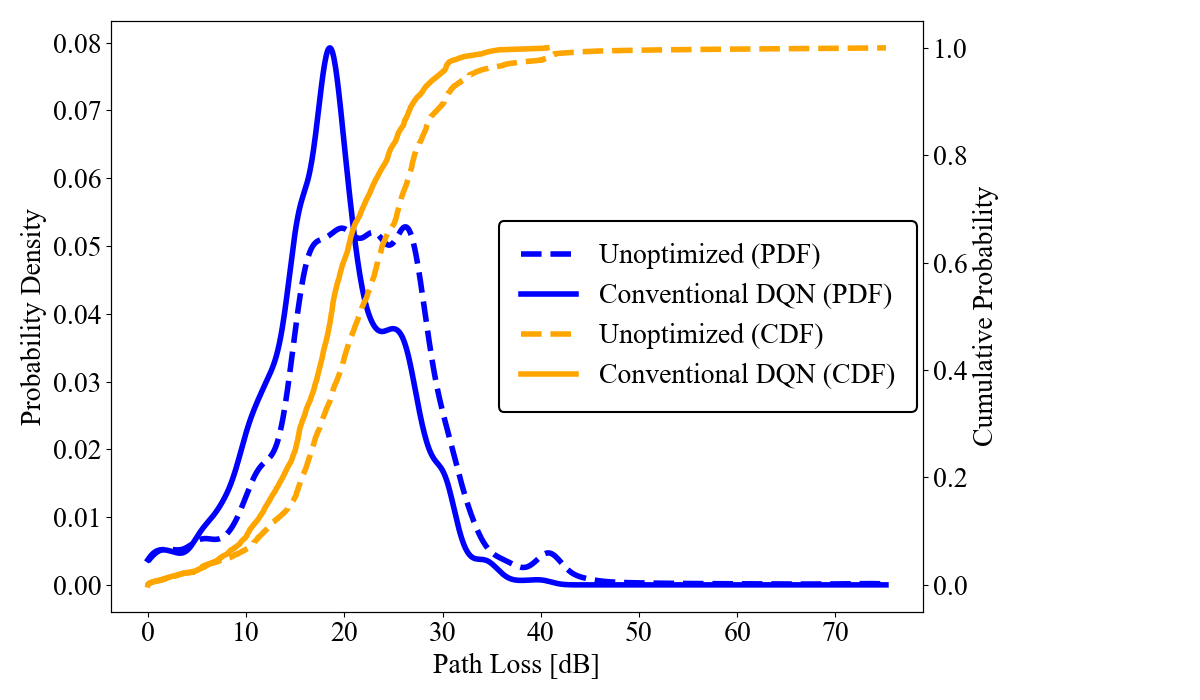} 
    \caption{PDF and CDF of DQN method.} 
    \label{figgl2} 
\end{figure}

\begin{figure}[htbp] 
    \centering 
    \includegraphics[width=0.5\textwidth]{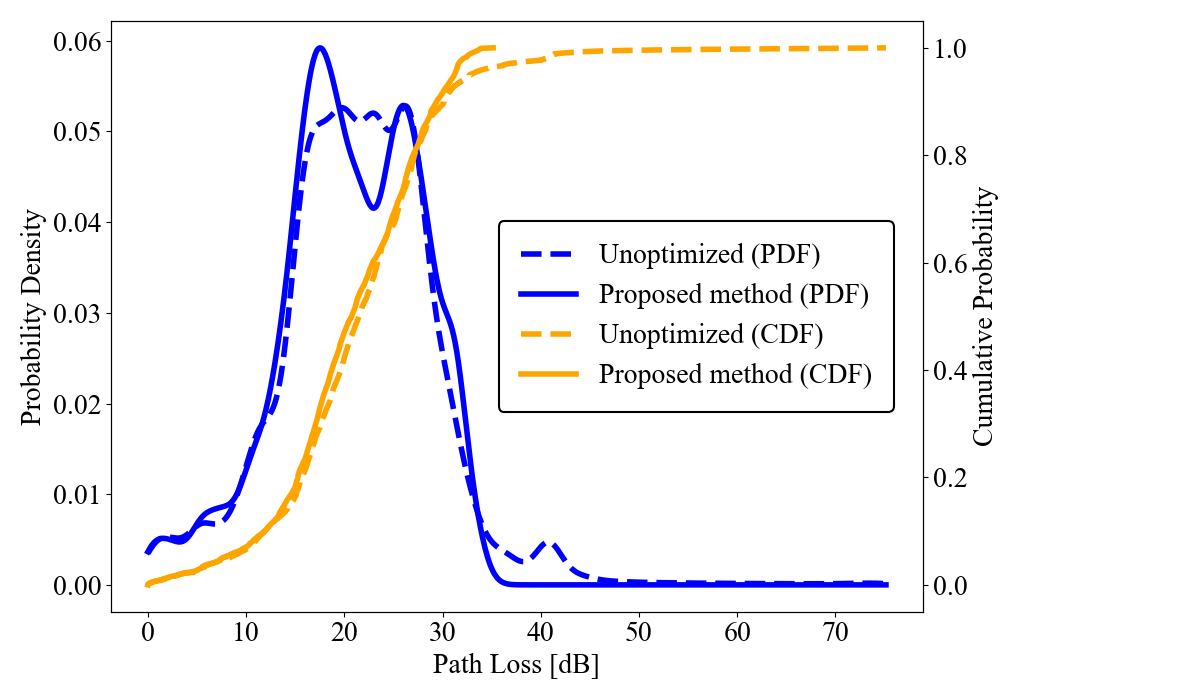} 
    \caption{PDF and CDF of Dueling DQN method.} 
    \label{figgl3} 
\end{figure}

\begin{algorithm}[!t]
\fontsize{11}{10}\selectfont
\caption{\fontsize{11}{10}\selectfont Dueling DQN for AP Deployment Optimization in Tunnels}
\label{alg:DuelingDQN_AP}
\begin{algorithmic}[1]
\STATE \textbf{Initialize:} Tunnel length $L=1500$, AP initial positions $\mathbf{X}_0=[0,500,1000]$, step size $\Delta$, state $s_t=[x_1^{(t)},x_2^{(t)},x_3^{(t)},g^{(t)}]$ 
\STATE \textbf{Initialize:} Dueling DQN parameters: shared layers $W_1\in\mathbb{R}^{64\times4}$, $W_2\in\mathbb{R}^{64\times64}$, value stream $V(s;\theta_v)$, advantage stream $A(s,a;\theta_a)$
\STATE \textbf{Initialize:} Replay memory $D$ (capacity $C=2000$), batch size $B=64$, $\gamma=0.995$, $\epsilon=1.0$, $\epsilon_{\text{decay}}=0.995$, $\epsilon_{\text{min}}=0.01$
\STATE \textbf{Initialize:} Dueling DQN network with coefficients $\theta$, target network $\theta^{-} \gets \theta$

\FOR{episode $=1$ \TO $N_{\text{episodes}}$}
    \STATE Reset environment: $\mathbf{X} \gets \mathbf{X}_0$, compute $g_0 = f(\mathbf{X})$, state $s_0 \gets [\mathbf{X}, g_0]$
    \WHILE{$g^{(t)} \geq 20$ \AND step $< \overline{N}_{\text{step}}$}
        \STATE Generate $r \sim \mathcal{U}(0,1)$
        \IF{$r < \epsilon$}
            \STATE Sample action index $a_{\text{idx}} \sim \mathcal{U}(0,26)$
        \ELSE
            \STATE Compute Q-values via Dueling DQN:
            
            $Q(s_t,a) =  \, V(s_t;\theta_v) + ( A(s_t,a;\theta_a) - \frac{1}{27}\sum_{a'} A(s_t,a';\theta_a))$
    
   \ENDIF
    \STATE Decode $a_{\text{idx}} = \underset{a}{\text{argmax}}\ Q(s_t,a)$

        \STATE Decode action vector: $(a_1,a_2,a_3) = \phi^{-1}(a_{\text{idx}})$ 
        
        \STATE Update AP positions:
        
        $x_i^{(t+1)} = \text{clip}\left(x_i^{(t)} + a_i \cdot \Delta,\ 0,\ L\right) \forall i=1,2,3$

        \STATE Compute $g^{(t+1)} = f(\mathbf{X}^{(t+1)})$, reward $R = -g^{(t+1)}$
        \STATE Store transition $(s_t, a_{\text{idx}}, R, s_{t+1})$ in $D$
        
        \IF{$|D| \geq B$}
    \STATE Sample minibatch $(s_j, a_j, R_j, s_{j+1})$
    \STATE Compute TD targets:
    \IF{$g^{(j+1)} < 20$}
    \STATE $y_j =R_j$
    \ELSE
    \STATE $y_j=R_j + \gamma \cdot \max\limits_{a'} \big[ Q(s_{j+1}, a';\theta^{-})\big] $
    \ENDIF

            \STATE Update $\theta$ via gradient descent on $\frac{1}{B}\sum (y_j - Q(s_j,a_j;\theta))^2$
        \ENDIF
        
        \IF{$\text{mod}(step, 100) == 0$}
            \STATE Update target network: $\theta^{-} \gets \theta$
        \ENDIF
        \STATE Decay $\epsilon \gets \max(\epsilon \cdot \epsilon_{\text{decay}}, \epsilon_{\text{min}})$
        \STATE $t \gets t+1$
    \ENDWHILE
\ENDFOR
\end{algorithmic}
\end{algorithm}

To align with the environmental model described in Section \uppercase\expandafter{\romannumeral3}, we specify the height of the AP (defined as the z axis) and the lateral distance from the tunnel wall (defined as the y axis) as 3 meters and 0.5 meters, respectively, thus restricting AP movement to the tunnel axis designated as the x axis. In addition, since the receivers employed on the train have a fixed height and distance from the tunnel wall as well, all receivers move on a one-dimensional curve along the x-axis. In that case, we only consider the path loss values on the curve of receivers and set APs on the one-dimensional curve along the x-axis, 3m high, and 0.5m from the tunnel wall. We discretize the curve where the receivers are located at intervals of 0.1 meters. Then, we use the PWE method to generate the path loss values of each discrete point when an AP is at different positions. Using the function \(\mathrm{PL}_i(\mathbf{X})\) mentioned in Section \uppercase\expandafter{\romannumeral2}, the path loss values can be calculated for scenarios with three APs deployed in the tunnel. For APs, we use a vertically polarized horn antenna with 7 dBi gain and 2.4 GHz operating frequency. For receivers, we use vertically polarized dipole-receiving antennas. The APs are placed at integer-meter positions on the curve where they are located. This means that during the training of deep reinforcement learning, the action step of an AP is also set to 1 meter. The three APs are initially placed at uniformly spaced intervals of 0m, 500m, and 1000m along the x-axis.

Additionally, a series of parameters of the functions in section \uppercase\expandafter{\romannumeral2} need to be set. The \(th_{i}\) defined in \autoref{th} is assigned a value of 30 to account for the path loss values generated by PWE. The \(\lambda_i\) defined in \autoref{f1} and \autoref{f2} is set to 10 to increase the intensity of penalties. The \(\alpha\) defined in \autoref{cost} is set at 0.5 to balance the impact of \(f_1\) and \(f_2\). In addition, we adjust \(\alpha\) to 0 and 1 to optimize when the cost function is only influenced by \(f_1\) or \(f_2\). Furthermore, the parameters of the DRL algorithm must be specified. \autoref{tab:DRL_parameters} illustrates the significant parameters.

\subsection{Performance Comparison of HJ, DQN, and Dueling DQN}
\autoref{tab:optimization_results} illustrates the optimization results of the three algorithms. The final AP placements and the final cost function have been listed. The cost function has been defined as \autoref{cost} in Section \uppercase\expandafter{\romannumeral2}. The final AP positions in \autoref{tab:optimization_results} show only the x-axis due to the fixed height configurations and the tunnel wall distance. In addition, the table shows the enhancements for the three methods compared to the initial AP positions. We can find that all of them successfully optimize the AP positions, since the cost function is reduced by them, but the cost function is maximally optimized by the Dueling DQN method while traditional DQN methods rank second. The HJ method performs the worst. Compared to HJ optimization, the traditional DQN method achieves a 7.5\% improvement. Meanwhile, the dueling DQN algorithm outperforms the traditional DQN with a 34.9\% improvement. \autoref{fig:optimization_results} shows the receive power of each receiver points under three distinguished methods. Analysis shows that the received power generally increases after optimization. In \autoref{figgl1}-\autoref{figgl3}, the probability density functions (PDF) and cumulative distribution functions (CDF) of path loss before and after optimization of the three methods comparison are shown. We can observe that all the statistical distribution profiles of the path loss move to the left after optimizations, which means that the path loss values are generally reduced.

\subsection{Results Comparison of Cost Function Weight Adjustments in Dueling DQN Algorithm}

\begin{figure*}[t]
    \centering
    \subfloat[]{%
        \includegraphics[width=0.32\textwidth]{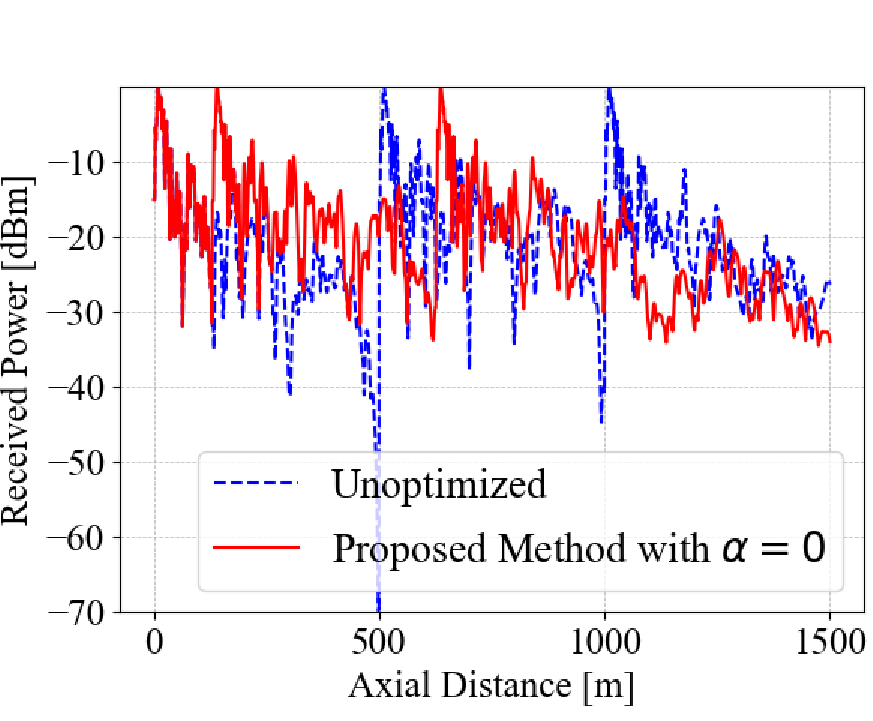}
    }\hspace{-1mm}
    \subfloat[]{%
        \includegraphics[width=0.32\textwidth]{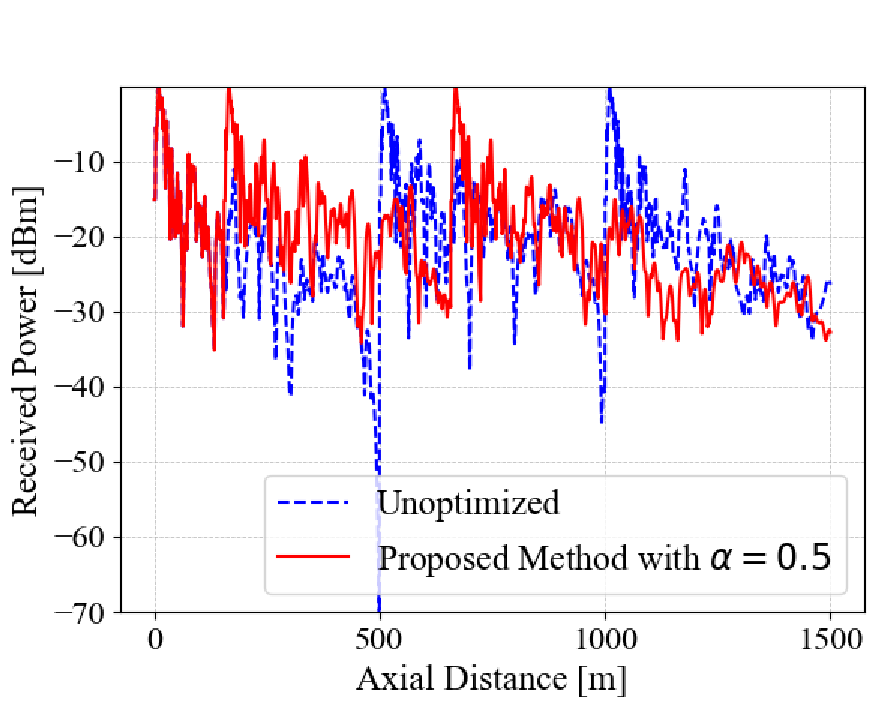}
    }\hspace{-1mm}
    \subfloat[]{%
        \includegraphics[width=0.32\textwidth]{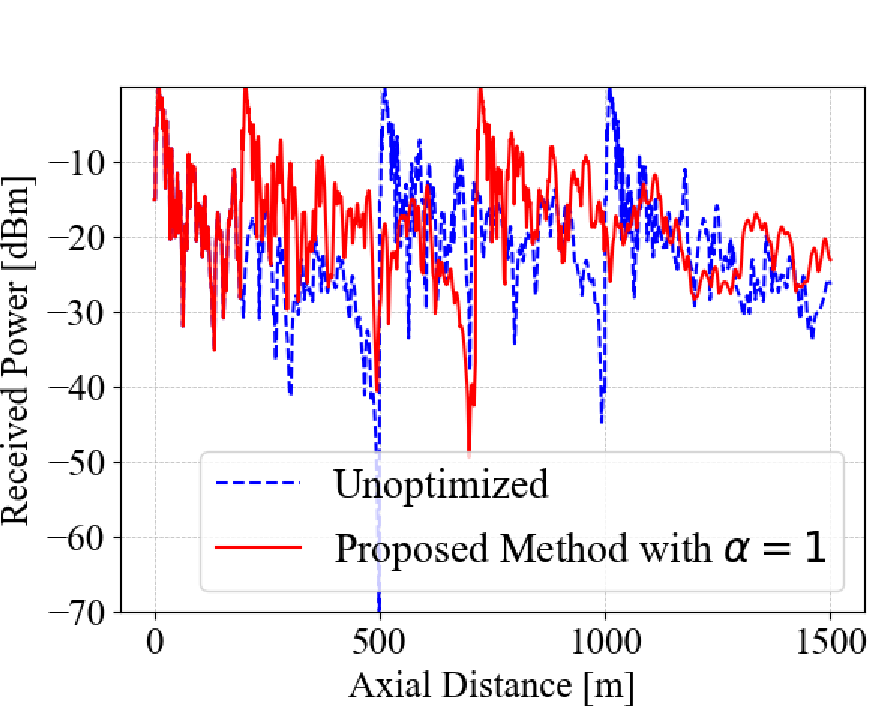}
    }
    \caption{Optimization results of the proposed Dueling DQN method under different values of $\alpha$:
(a) $\alpha = 0$;
(b) $\alpha = 0.5$;
(c) $\alpha = 1$.}
    \label{fig:optimization_results_alpha}
\end{figure*}

We change the weight parameter \(\alpha\) in the cost function defined as \autoref{cost} to find its impact on the Dueling DQN optimization results. The results are displayed in \autoref{tab:parameter_adjust}. We also present \autoref{fig:optimization_results_alpha} to demonstrate the receive power values at each receiver point in different \(\alpha\). When \(\alpha\) is set to 0, the optimization focuses more on the receiver points with high path loss values. When \(\alpha\) is set to 1, the optimization only considers the impact of the average path loss. When \(\alpha\) is set to 0.5, both the average path loss and the point with the worst path loss are considered. In that case, the weight parameter \(\alpha\) can be chosen based on different user requirements.

\subsection{Performance of cGAN-based Path Loss Generation}

To evaluate the quality of the generated path loss (PL) maps, we use two standard error metrics: Mean Squared Error (MSE) and Mean Absolute Error (MAE). Both are computed over the physically valid, forward-direction region of the tunnel (after the AP location), defined by the binary mask $m$:
\begin{equation}
\mathrm{MSE} = \frac{1}{N} \sum_{i} m_i (y_i - \hat{y}_i)^2
\end{equation}
\begin{equation}
\mathrm{MAE} = \frac{1}{N} \sum_{i} m_i |y_i - \hat{y}_i|
\end{equation}
where $y_i$ is the PWE-simulated reference path loss and $\hat{y}_i$ is the cGAN-generated prediction.

\autoref{tab:error_stats} summarizes the aggregated error metrics over all AP positions in the tunnel. The results demonstrate that the proposed cGAN model achieves low mean and 90th-percentile errors across the entire tunnel length.

\begin{table}[h]
\small
\centering
\caption{Overall Error Metrics (Grouped by Meter)}
\begin{tabular}{lcc}
\hline
Metric & Mean $\pm$ Std & 90th Percentile \\
\hline
MSE (dB$^2$) & 0.08080 $\pm$ 2.74177 & 0.02325 \\
MAE (dB)     & 0.05676 $\pm$ 0.26662 & 0.09459 \\
\hline
\end{tabular}
\label{tab:error_stats}
\end{table}

These values indicate that even when generating high-resolution PL maps for unseen AP positions, the model maintains low average prediction error. The 90th-percentile results further confirm the reliability of the model under most deployment scenarios.

\begin{figure}[h]
    \centering
    \includegraphics[width=0.41\textwidth]{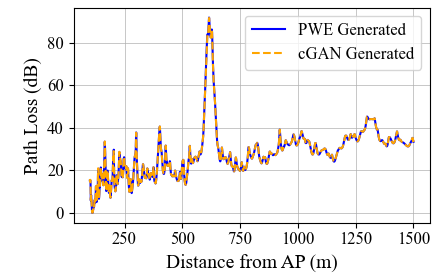}
    \caption{Path loss at AP position 100\,m: PWE-simulated vs cGAN-generated.}
    \label{fig:ap100}
\end{figure}

\begin{figure}[h]
    \centering
    \includegraphics[width=0.41\textwidth]{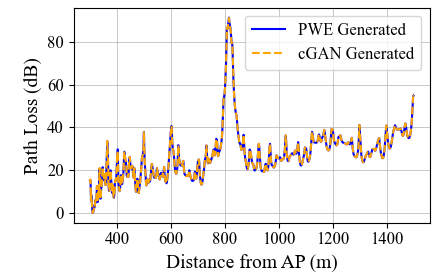}
    \caption{Path loss at AP position 500\,m: PWE-simulated vs cGAN-generated.}
    \label{fig:ap500}
\end{figure}

\autoref{fig:ap100} and \autoref{fig:ap500} illustrate qualitative results for APs deployed at 100\,m and 500\,m, respectively. Each plot compares the PWE-simulated reference PL curve with the cGAN-generated prediction. In both cases, the cGAN successfully captures the complex fading patterns and overall signal attenuation trends, closely matching the reference data in the valid forward region.

\begin{figure}[h]
    \centering
    \includegraphics[width=0.45\textwidth]{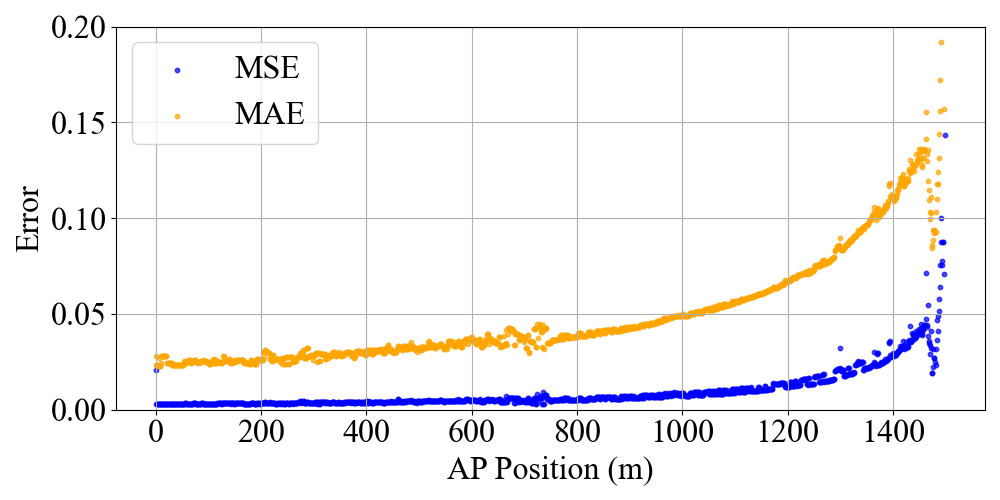}
    \caption{Scatter plot of MSE and MAE vs AP position along the tunnel.}
    \label{fig:mse_mae_scatter}
\end{figure}

\autoref{fig:mse_mae_scatter} shows the distribution of MSE and MAE as a function of AP position along the tunnel. Each point represents the grouped error (by meter), providing insight into how prediction quality varies with AP location. Errors generally remain low for most AP positions, with a slight increase toward the far end of the tunnel. This is likely due to reduced training data density in those regions, highlighting potential benefits from targeted data augmentation.

Overall, the cGAN model demonstrates strong capability in reproducing high-resolution PL distributions from limited PWE simulation data. Its ability to preserve fine-scale fading structures and attenuation trends ensures that the generated PL maps can serve as reliable surrogates for computationally expensive full-wave simulations in subsequent optimization tasks.

\section{Conclusion}
This paper proposes a novel physics-driven and data-augmented framework for optimal access point deployment in railway tunnels within the communication-based train control system. Unlike existing AP placement optimization methods, which often rely solely on heuristic search or purely data-driven models and thus suffer from local optima or limited generalization, the proposed approach integrates high-fidelity channel modeling using the parabolic wave equation with a conditional generative adversarial network for data augmentation, and a dueling deep Q-network for policy learning.

In the proposed framework, a limited number of PWE simulations are first conducted to generate accurate path loss maps under different AP placements. These high-fidelity samples are then used to train the cGAN, which learns to synthesize realistic PL maps for untested AP positions, substantially reducing the computational cost of exhaustive electromagnetic simulations. The augmented PL dataset is subsequently used to define a CBTC-specific reinforcement learning environment, enabling the Dueling DQN agent to explore and optimize AP configurations efficiently. Comparative experiments with Hooke and Jeeves optimization and standard DQN confirm that our method achieves superior average and worst-case received power performance, with the additional benefit of reduced simulation requirements.

The framework also supports flexible optimization objectives by tuning the parameter \(\alpha\) in the cost function: a higher \(\alpha\) emphasizes minimizing the average PL, whereas a lower value prioritizes worst-case PL improvement. This adaptability allows the system to be customized for different CBTC operational requirements while maintaining superior performance over baseline methods.

Future research will focus on integrating more advanced deep reinforcement learning algorithms, such as rainbow DQN, which combines enhancements including double DQN, Prioritized Experience Replay, Noisy Nets, and categorical DQN. These techniques could further improve exploration-exploitation balance, adapt policies dynamically to spatially varying tunnel conditions, and accelerate convergence by prioritizing high-impact samples (e.g., from severe attenuation zones). Such improvements would make the proposed hybrid framework even more scalable, robust, and deployment-ready for complex tunnel environments.

\ifCLASSOPTIONcaptionsoff
\newpage
\fi


\end{document}